\documentclass[%
 reprint,
 amsmath,amssymb,
 aps,
]{revtex4-2}

\usepackage{graphicx}
\usepackage{amsmath,amssymb}
\usepackage{bm}
\usepackage[T1]{fontenc}
\usepackage{siunitx}
\DeclareSIUnit{\keV}{keV}
\DeclareSIUnit{\arcsecond}{\ensuremath{''}}

\newcommand{\kx}{k_x}
\newcommand{\Jk}{J(\kx)}
\newcommand{\Wxk}{W(x,\kx)}
\newcommand{\seff}{\sigma_e}

\begin{document}

\preprint{APS/123-QED}

\title{Direct Observation of X-ray Double-Slit Interference in Momentum Space}

\author{Fugui Yang }
 \email{yangfg@ihep.ac.cn}
 \affiliation{Institute of High Energy Physics, Chinese Academy of Sciences, Beijing 100049, China}
 
\author{Xiaoxiao Liang}
 \affiliation{Institute of High Energy Physics, Chinese Academy of Sciences, Beijing 100049, China}

\author{Tianchong Zhang}
 \affiliation{Institute of High Energy Physics, Chinese Academy of Sciences, Beijing 100049, China}

\author{Xiaowei Zhang }
 \email{zhangxw@ihep.ac.cn} 
 \affiliation{Institute of High Energy Physics, Chinese Academy of Sciences, Beijing 100049, China}

\date{\today}

\begin{abstract}
Young’s double-slit experiment is conventionally deemed a spatial phenomenon emerging from free-space transport. In this Letter, we invert this perspective to demonstrate that Young's interference can be accessed directly as a pure momentum-space observable. Using a perfect-crystal diffraction to project the field's reciprocal-space profile immediately downstream of the aperture, we resolve the complete hard X-ray double-slit fringe structure without any propagation arm, focusing optics, or imaging detector. This direct capture of the field's invariant momentum marginal establishes a compact, lensless, and propagation-free approach to coherence diagnostics, proving that the fundamental physics of wave interference can be detached from real-space propagation.

\end{abstract}

\maketitle

Young's double-slit experiment is among the most beautiful and iconic demonstrations of wave interference ~\cite{Crease2002}  and has profoundly shaped modern physics. Since its realization with visible light in 1804 ~\cite{Young1804}, the same physical principle has been repeatedly realized across disparate physical systems, including photons, electrons ~\cite{Jonsson1961}, neutrons~\cite{Zeilinger1988}, and atoms etc.~\cite{Carnal1991,Keith1988},  establishing interference as a universal manifestation of quantum coherence across scales.

Its extension to hard X-rays, by contrast, came with a striking delay---X-ray interference was first inferred from the Bonse--Hart interferometer~\cite{BonseHart1965}  by Laue diffractions in a monolithic silicon block. Young double-slit coherence measurements with synchrotron X-rays were pioneered in the soft-X-ray regime~\cite{Takayama1998} and extended to the hard-X-ray band with third-generation sources~\cite{Leitenberger2001}. This delay does not reflect any fundamental limitation of the underlying physics, but rather arises from the combined constraints of ultra-short wavelength and experimental implementation.  A real-space transcription of Young's geometry demands micron-scale apertures of extreme aspect ratio or long transport distance, a high-brilliant source, and high-resolution imaging area detection. The hard X-ray case makes the cost explicit(110 m in a recent single-photon realization~\cite{Gureyev2024}).  These unconventional requirements reflect not a conceptual limitation of interference, but the geometric encoding strategy employed in real-space implementations. 

A more general strategy is to relax the constraints of spatial encoding and reconsider the underlying measurement basis. In this Letter, we demonstrate that the essential interference physics can be directly accessed through a momentum-space formulation, which we term the Momentum-Space Interferometer (MSI). While conventional real-space measurements conflate the generation of interference information ("born at the slit") with its subsequent transport ("transported by propagation"), our momentum-representation approach proves that the complete interference structure is already fully encoded within the momentum boundary immediately behind the slits—requiring neither free-space propagation nor lens-based focusing. This paradigm shift offers a fundamentally more direct, compact, and calibration-free route to observe and utilize hard X-ray interference.

\begin{figure*}[t]
\includegraphics[width=\linewidth]{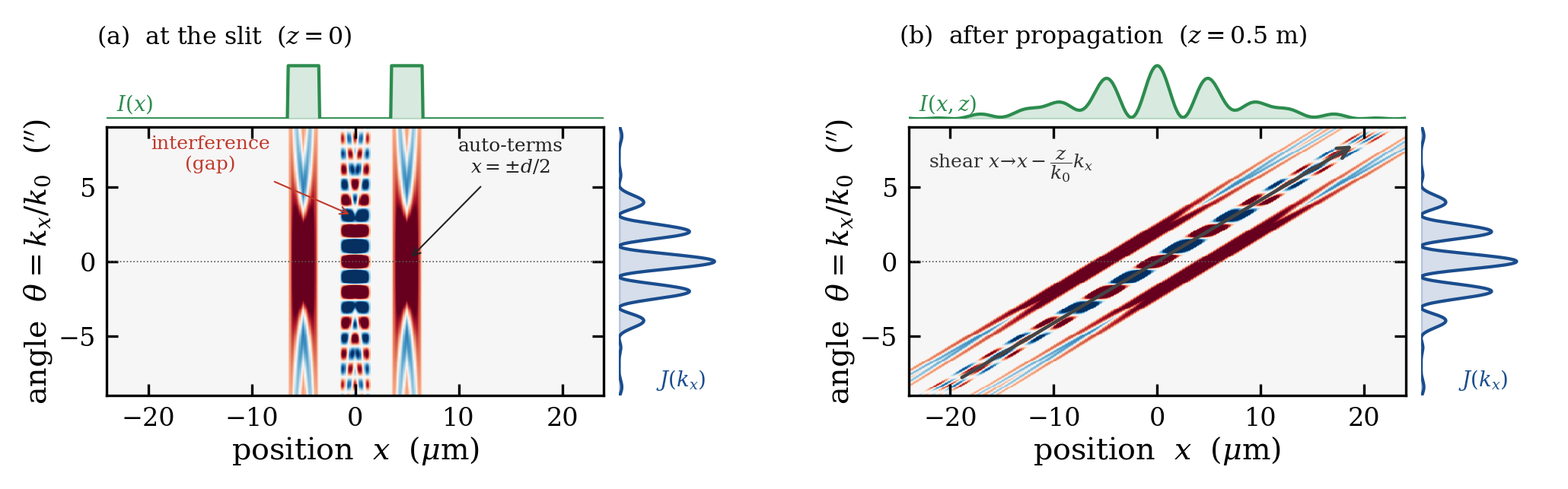}
\caption{Phase-space (Wigner) picture of the double slit. (a) At the slit ($z=0$),
the Wigner density $\Wxk$ has two auto-terms at the geometric slit positions
$x=\pm d/2$ and a $\cos(\kx d)$-oscillating interference term in the empty gap at
$x=0$. The position marginal $I(x)$ (top) is just two slits, whereas the momentum
marginal $\Jk$ (right) is already the full Young pattern of angular period
$\lambda/d$---the quantity the analyzer records. (b) After free propagation
($z=\SI{0.5}{\meter}$) the density is sheared, $x\to x-(z/k_0)\kx$
[Eq.~\eqref{eq:shear}], yet the momentum marginal $\Jk$ (right) is unchanged
(propagation invariant), while the position marginal $I(x,z)$ (top) evolves toward
the far-field Young fringe pattern. The fringe information lives in $\Jk$ at the
slit and is merely transported onto position by propagation.}
\label{fig:concept}
\end{figure*}

\emph{Phase-space picture.}---A monochromatic field is described, at any plane, by
its Wigner phase-space density~\cite{Wigner1932,Bastiaans1978,Walther1968,Raymer1994}
\begin{equation}
\Wxk=\int E\!\Big(x+\tfrac{x'}{2}\Big)E^*\!\Big(x-\tfrac{x'}{2}\Big)\,
e^{-i\kx x'}\,dx',
\label{eq:wigner}
\end{equation}
the \emph{master object} of the measurement, a real-valued (and, for a wave field,
partly negative) joint density on transverse position $x$ and momentum
$\kx=k_0\theta$, which for a partially coherent field generalizes by
$E(x_1)E^*(x_2)\!\to\!\Gamma(x_1,x_2)$ to the radiometric phase-space
brightness~\cite{Bastiaans1986,MandelWolf1995}. Every linear intensity measurement
is one phase-space kernel read against this single object,
\begin{equation}
m[K]=\langle K,W\rangle=\iint K(x,\kx)\,\Wxk\,\frac{dx\,d\kx}{2\pi},
\label{eq:mmap}
\end{equation}
the method residing entirely in the choice of $K$. Its two canonical kernels, corresponding to a planar detector and an angle-selective analyzer, return the measurable spatial and momentum marginals,
\begin{subequations}\label{eq:marginals}
\begin{align}
\int \Wxk\,\frac{d\kx}{2\pi}&=|E(x)|^2\equiv I(x),\label{eq:marg-x}\\
\int \Wxk\,dx&=|A(\kx)|^2\equiv \Jk.\label{eq:marg-k}
\end{align}
\end{subequations}
where $A(\kx)=\int E(x)\,e^{-i\kx x}\,dx$ is the Fourier transform of the field.  The momentum-marginal identity $\Jk=|A(\kx)|^2$ is purely algebraic. This algebraic identity holds at any fixed transverse plane, independent of propagation dynamics.

For a coherent double slit of separation $d$ and width $w$, the Wigner density has
a vivid three-term structure [Fig.~\ref{fig:concept}(a)],
\begin{equation}
\begin{split}
\Wxk={}&W_0\!\Big(x-\tfrac{d}{2},\kx\Big)+W_0\!\Big(x+\tfrac{d}{2},\kx\Big)\\
       &{}+2\,W_0(x,\kx)\cos(\kx d),
\end{split}
\label{eq:wigner3}
\end{equation}
where $W_0(x,\kx)$ is the single-slit Wigner function---two ``auto-term'' lobes at the
geometric slit positions $x=\pm d/2$, and a third (cross) term in the \emph{empty
gap} at $x=0$ that oscillates along the momentum axis as $\cos(\kx d)$ and carries
the fringe information~\cite{Bastiaans1986}. Projecting onto $k_x$ yields the standard Young pattern $J(k_x) = \text{sinc}^2(wk_x/2\pi)[1+\cos(k_x d)]$, exhibiting a cosine fringe of angular period $\lambda/d$ under a single-slit envelope of width $\lambda/w$, present \emph{immediately} at the slit; projected onto $x$ at $z=0$
the same gap term integrates to zero, so the real-space intensity at the mask is
just two slits, with no fringes. This mean that the fringe information is \emph{born} in the momentum projection, not produced by propagation.

By the optical Noether theorem, the translational symmetry of the free-space Hamiltonian $H(k_x)$ ensures the exact conservation of transverse momentum. The Wigner evolution equation $\partial_z W = \{H, W\}_{\text{PB}}$ then satisfies $\partial_z J(k_x) = 0$ for all $z$, strictly holding to all orders without paraxial approximations~\cite{Bastiaans1986}. Reading $\Jk$ directly---\emph{momentum-space interferometry}
(MSI)---measures this conserved quantity at any distance behind the slit. For the traditional Young's interference, the position pattern, meanwhile, evolves as a rigid shear~\cite{Bastiaans1979},
\begin{equation}
W_z(x,\kx)=W_0\!\Big(x-\tfrac{z}{k_0}\kx,\;\kx\Big),
\label{eq:shear}
\end{equation}
carrying $\Jk$ onto position at large $z$, with $I(x,z)\to(k_0/z)\Jk|_{\kx=k_0x/z}$, so that a position detector at large $z$ becomes a $\kx$-analyzer; the far-field experiment accesses $\Jk$ only incidentally by this shear. Equation (5) reveals an uncompromising truth: free-space propagation acts merely as a passive phase-space shear ($x \rightarrow x - \frac{z}{k_0}k_x$). It creates no new interference information; it merely translates an existing momentum-space invariant onto the position axis. The textbook Young's fringes observed on a distant screen are nothing but a real-space ghost of this invariant momentum marginal. Bypassing this detour, our Momentum-Space Interferometer (MSI) captures this prime object at its birth.

For hard X-ray, a perfect crystal analyzer acts as a momentum projector or realizes the momentum kernel $K=H$ with a \emph{computed}, not calibrated, response. ~\cite{Darwin1914,BattermanCole1964,Authier2001}; it is therefore a narrow
band-pass $H(\kx-k_\omega)$ in transverse momentum, of width $\delta =k_0\Delta_D$ within the Darwin width $\Delta_D$. Scanning $\omega$ sweeps the pass-band across $\kx$, and the measurement map~\eqref{eq:mmap} reduces to
\begin{equation}
R(\omega)=\langle K_\omega,W\rangle=(J\!\ast\!H)(\omega),
\label{eq:rock}
\end{equation}
the momentum marginal $\Jk$ smoothed by the known, lattice-defined instrument
function $H$. The rocking curve \emph{is} the two-slit interference pattern, read in
transverse momentum, with the crystal playing the role that an observation screen
plays in the conventional experiment---but in the conjugate basis.

\begin{figure}[t]
\includegraphics[width=\linewidth]{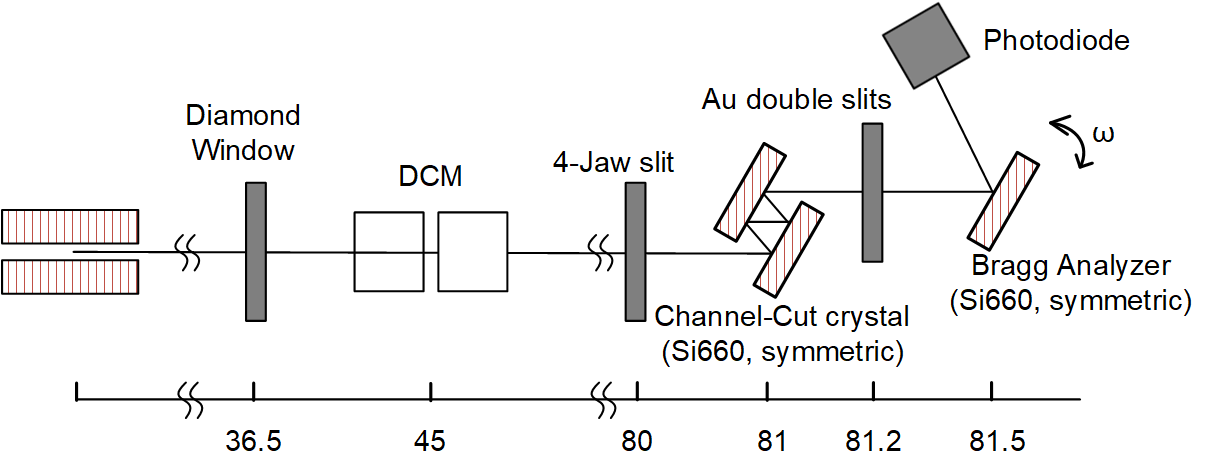}
\caption{Experimental configuration (schematic top view). The interferometer consists of a high-resolution monochromator using a channel-cut Si(660) crystal, an Au double slit,  precision four-jaw slit, a Si(660) Bragg analyzer, and an photodiode (PD). For the
duality cross-check the analyzer is removed and a 2-D imaging detector at
$D=\SI{8}{\meter}$ (not shown) records the real-space far-field pattern of the same
slit.}
\label{fig:config}
\end{figure}

\emph{Experiment.}---The measurement was performed at ID05 - a hard-X-ray
undulator beamline of the High Energy Photon Source (HEPS)~\cite{HEPS2018} at a
photon energy of \SI{12.4}{\keV}. The optical layout is shown in Fig.~\ref{fig:config}. A Si(111) double-crystal monochromator(DCM) is followed by a  Si(660) channel-cut high resolution monochromator arranged non-dispersively $(+,-)$ with respect to a downstream Si(660) Bragg analyzer of the same reflection. The double slit (nominal separation $d=\SI{10}{\micro\meter}$, opening $w=\SI{3}{\micro\meter}$) is a high-aspect-ratio gold absorber of thickness $t\approx\SI{30}{\micro\meter}$ on a low-$Z$ membrane, fabricated by deep X-ray lithography and electroforming. The slit is mounted with a small overall tilt $\alpha$ of its walls relative to the beam (a controlled degree of freedom used below). The symmetric Si(660) analyzer ($\theta_B=\ang{51.36}$, Darwin width \SI{0.66}{\arcsecond}) is rocked through $\omega$ (\SI{400}{} encoder pulses per arcsecond) while a photodiode(PD) integrates the reflected flux. For the duality cross-check, a camera (pixel pitch \SI{4.6}{\micro\meter}) at $L\approx\SI{8}{\meter}$ records the real-space far-field pattern of the same slit. Throughout, the instrument function $H$ is determined in the same scan configuration from a no-slit run.

\begin{figure}[t]
\includegraphics[width=\linewidth]{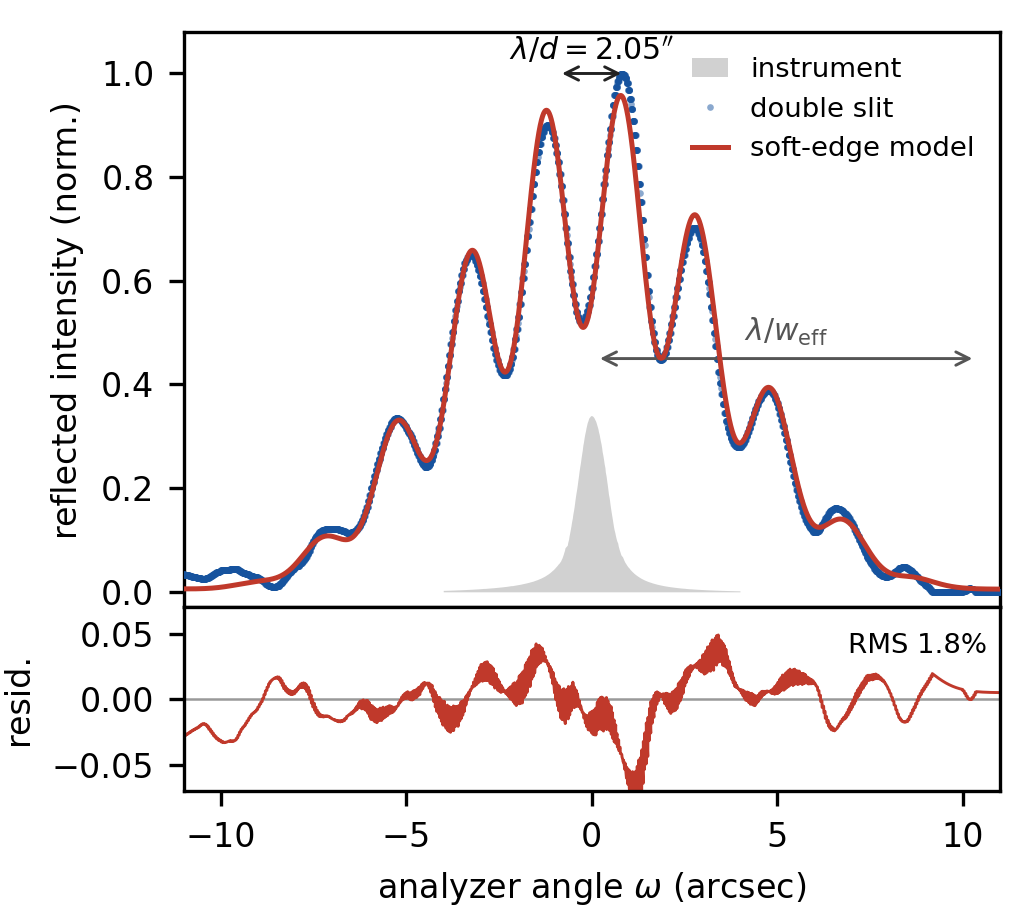}
\caption{The double-slit interference pattern observed in momentum space. Measured
rocking curve of the double slit (points) and the soft-edge forward model (line),
convolved with the independently measured instrument function (shaded). The fringe
period is $\lambda/d=\SI{2.05}{\arcsecond}$ ($d=\SI{10.1}{\micro\meter}$); the
single-slit envelope sets the angular scale $\lambda/w_{\rm eff}$. }
\label{fig:rocking}
\end{figure}

Figure~\ref{fig:rocking} shows the central result. The rocking curve of the double slit (points) is the Young pattern, recorded in transverse momentum. A cosine fringe of measured angular period $\lambda/d=\SI{2.05}{\arcsecond}$ sits under a single-slit envelope, returning the fringe separation $d=\SI{10.1}{\micro\meter}$ directly from the angular period, independent of the instrument resolution. The instrument function, measured on a no-slit run in the same configuration (full width $\approx\SI{0.96}{\arcsecond}$, shaded), decomposes into the analyzer Darwin width and the channel-cut energy width in quadrature. The informative deviation from the ideal Young pattern lies not in the fringes but in the \emph{envelope}. A hard-edged slit would give the $\mathrm{sinc}^2(w\kx/2\pi)$ single-slit envelope, yet the data prefer a more peaked, side-lobe-suppressed one (a hard edge under-predicts the central fringe by $3$--$8\%$). We therefore model the single slit as an \emph{apodized} aperture, $a(x)=\mathrm{rect}(x/w_{\rm eff})\!\ast\!G_{\seff}(x)$---physically expected for the deep, tilted gold walls, whose graded absorber path tapers the transmission over a sub-micron edge. A convolved-domain fit reproduces the rocking curve to $\approx\SI{1.8}{\percent}$ rms with edge width $\seff\approx\SI{0.3}{\micro\meter}$, effective width $w_{\rm eff}\approx\SI{2.0}{\micro\meter}$, and visibility $V\simeq0.94$ (the high $V$ reflecting the coherent illumination, $\xi\gg d$). The model, the fit pipeline, and a controlled-tilt series that measures the absorber thickness $t\approx\SI{30}{\micro\meter}$ through $w_{\rm eff}=w_{\rm nom}-t\tan\alpha$ are detailed in the Supplemental Material-~S1.

\begin{figure}[t]
\includegraphics[width=\linewidth]{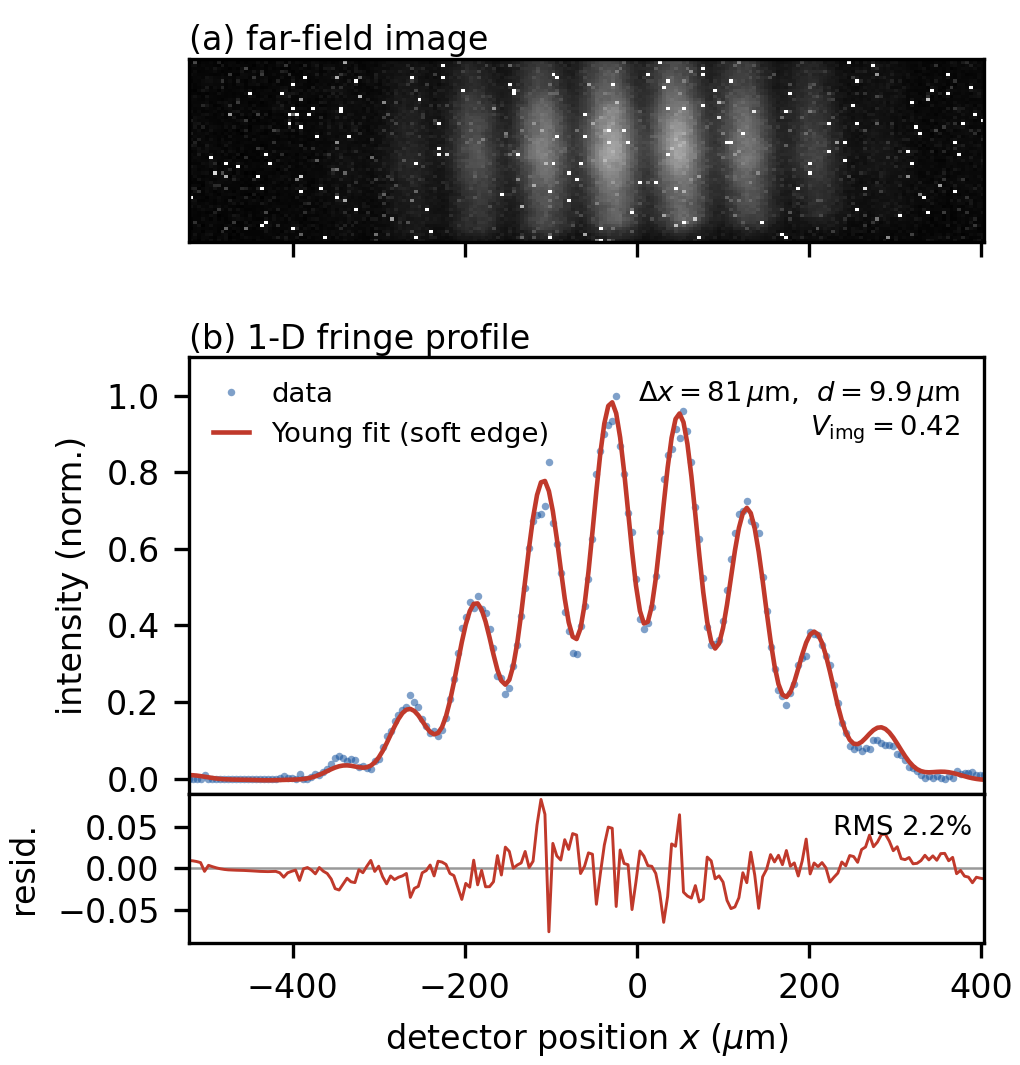}
\caption{Real-space far-field interference of the same double slit (duality
cross-check). (a) Direct 2-D detector image at $L\approx\SI{8}{\meter}$; the Young
fringes run horizontally (slits separated vertically). (b) Background-subtracted
one-dimensional fringe profile (points) with the soft-edge two-slit fit (line), of
period $\Delta x=\SI{81}{\micro\meter}$. The reduced image contrast $V_{\rm img}=0.42$
calibrates the detector point-spread function.}
\label{fig:interference}
\end{figure}

We also record the slit the conventional way. The 2-D detector image [Fig.~\ref{fig:interference}(a)] resolves $\sim11$ Young fringes; the background-subtracted one-dimensional profile [Fig.~\ref{fig:interference}(b)] is fit by the same soft-edge two-slit model.  This reproduces the textbook experiment but exacts the position-representation cost in full---a multi-metre propagation arm, and a contrast $V_{\rm img}=0.42$ set by the detector point-spread function rather than the slit. Through the duality relation $\Delta x/L=\lambda/d$ this is the same fringe content as the momentum-space scan. With the camera distance $L\approx\SI{8}{\meter}$ the real-space period returns $d=\SI{9.9}{\micro\meter}$, and the calibration-free ratio of envelope to fringe scale gives $d/w_{\rm eff}=5.1$, both matching the momentum-space values ($d=\SI{10.1}{\micro\meter}$, $d/w_{\rm eff}=5.0$) to within a few percent. 

\emph{Discussion.}---In the phase-space picture every transverse measurement reads
one of three projections of the master object $W$---the position marginal $I(x)$
($x$--$x$, imaging), the momentum marginal $\Jk$ ($p$--$p$, far field or analyzer),
or the joint density. The MSI result places Young's double slit unambiguously in the
\emph{momentum} ($p$--$p$) class, since at the mask the fringe pattern lives only in
$\Jk$---the position marginal there is just two slits---and $\Jk$ is a propagation
invariant. The conventional far-field experiment accesses this
momentum quantity only \emph{incidentally}, because free propagation shears the angular
content onto position, so that at large $z$ a position detector has effectively
become a $\kx$-analyzer, and the multi-metre arm and the extreme detector and optic
resolution are the price of reading a momentum quantity with a position instrument.
The analyzer crystal removes the detour and reads the same invariant directly at the
slit. A perfect crystal thus provides an \emph{ab initio} momentum-space projector for hard X-rays---selecting transverse momenta at the Darwin-width resolution, a reproducible property of the lattice that carries no manufactured-optic tolerance budget and needs no far field. Because the momentum marginal generalizes to the phase-space brightness of a partially coherent field~\cite{Bastiaans1986,Walther1968}, the same instrument is also a direct, propagation-free probe of transverse coherence---complementary to propagated-fringe methods~\cite{Pfeiffer2005,Yabashi2001,Vartanyants2011,Gutt2012,Singer2013}.

Beyond reading the fringe, the momentum marginal carries the full
single-slit transfer function, so here it resolves the apodized edge profile of the
aperture and, through the controlled-tilt projection, the absorber thickness---a
metrology of the diffracting structure that the bare far-field fringe does not
provide. For hard X-ray, the high height -width ratio makes the soft-edge envelope neccessary, not peculiar to our device. The same effective-width reduction and a mild fringe asymmetry recur in other hard-X-ray double-slit experiments~\cite{Gureyev2024,Castle2025}, where they are noted only in passing---ascribed to a non-uniform background or to unequal slit widths---without a quantitative model. Neither attribution suffices, because the far-field intensity of any real (amplitude-only) aperture, unequal slits included, is even in $\kx$ and so cannot by itself produce a left--right asymmetry, which requires an odd slit-to-slit phase $\varphi$. Because the rocking curve \emph{is} the momentum marginal, the apodized envelope and this residual $\varphi$ are separated cleanly rather than entangled with a propagation distance, turning a recurring and previously unmodelled nuisance into a quantitative diagnostic of the aperture.

In summary, we have performed the first direct measurement of Young’s double-slit interference as a momentum-space observable in the hard-X-ray regime. The observed pattern corresponds to the transverse momentum distribution of the transmitted field, which is fully established immediately behind the slits and remains invariant under free-space propagation. By using a perfect-crystal analyzer as a reciprocal-space projector, the interference structure is accessed directly without a screen, a far field, or focusing optics. Independent real-space measurements of the same aperture confirm the consistency between the real-space and momentum-space representations. Our results establish perfect-crystal diffraction as a compact, lensless, and lattice-referenced platform for momentum-space interferometry and open a direct route to transverse-coherence measurements at diffraction-limited X-ray sources.

\begin{acknowledgments}
This work is supported by the Strategic Priority Research Program of the Chinese Academy of Sciences (Grant No. XDA0520502), National Key Research and Development Program of China (Grant No. 2024YFF0617000). We thank the ID05-Low-Dimension Structure Probe Beamline of High Energy Photon Source (https://cstr.cn/31138.02.HEPS.ID05) for providing technical support and assistance in diffraction data collection.
\end{acknowledgments}

\bibliographystyle{apsrev4-2}
\bibliography{refs}

\end{document}


\title{Supplemental Material:\\ Observation of X-ray Double-Slit Interference in Momentum Space}
\author{Author One}
\author{Author Two}
\affiliation{Affiliation, City, Country}
\date{\today}
\maketitle

This Supplemental Material gives the soft-edge (apodized) aperture model used in the
main text and its verification: the convolved-domain data-analysis pipeline that
extracts the slit parameters from the momentum-space rocking curve, and the
controlled-tilt series that measures the gold absorber thickness. Figure numbers
without an ``S'' refer to the main text.

\section{Data-analysis pipeline for the rocking curve (Fig.~3)}
\label{sec:pipeline}

The double-slit rocking curve is analysed entirely in the \emph{convolved domain}: the
model is convolved with the independently measured instrument function and compared with
the raw data, rather than deconvolving the data. Richardson--Lucy deconvolution of the
rocking curve over-sharpens the sinusoidal fringes into spikes (a round-trip test on a
synthetic sinusoid reproduces $\approx99\%$ of the apparent ``non-sinusoidality,'' i.e.\ it
is an artifact); the convolved-domain fit is artifact-free. No symmetrization is applied at
any stage.

\paragraph{Instrument function $H$.}
A no-slit rocking curve recorded in the same configuration (si\_660\_18) is floor-subtracted
(median of the signal-free wings), lightly smoothed, peak-centred, interpolated onto a
uniform $\SI{0.005}{\arcsecond}$ grid, and area-normalised ($\int H=1$). Its \emph{raw,
intrinsically asymmetric} shape is used as $H$ (full width $\approx\SI{0.96}{\arcsecond}$);
the thick gold blocks the divergent beam component, so the open-beam curve---not a
broader gold-plus-guard run---is the appropriate kernel.

\paragraph{Preprocessing of the double-slit curve.}
The curve is interpolated onto the same grid, floor-subtracted, normalised to its smoothed
peak, and peak-centred.

\paragraph{Stage 1 -- soft-edge envelope.}
Fringe-peak positions are located (\texttt{scipy.find\_peaks} with a one-period minimum
spacing). The single-slit amplitude is modelled as an apodized (super-Gaussian) aperture,
$a(x)=\mathrm{rect}(x/w_{\rm eff})\ast G_{\sigma_e}(x)$, with envelope $|\mathcal{F}\{a\}|^2$
---equivalently the hard-edge amplitude multiplied by
$g(\kx)=\exp(-\tfrac12\sigma_e^2\kx^2)$, which suppresses the slowly decaying
$\mathrm{sinc}^2$ side-lobes and raises the central lobe. Its amplitude, flat-top width
$w_{\rm eff}$, edge width $\sigma_e$, and centre $\omega_0$ are
least-squares fitted to the fringe-peak \emph{heights}, giving
$w_{\rm eff}\approx\SI{1.97}{\micro\meter}$ and $\sigma_e\approx\SI{0.30}{\micro\meter}$. The
resulting apodized aperture, and its far-field envelope with the side-lobes suppressed so as
to match the measured fringe-peak heights, are shown in Fig.~\ref{fig:soft}(a); a hard-edge
$\mathrm{sinc}^2$ would instead under-predict the central fringe by $3$--$8\%$.

\paragraph{Stage 2 -- fringe fit.}
With the envelope fixed, the full two-slit model
$I(\omega)=A\,[\,|a(\kx)|^2(1+V\cos(\kx d+\varphi))\,]\ast H+B$ is fitted; the pitch $d$ is
scanned on a grid (the $\cos(\kx d)$ term is multimodal in $d$) and $(A,V,\varphi,B)$ are
least-squares optimised at each $d$, retaining the lowest-rms solution. The result is
$d=\SI{10.08}{\micro\meter}$ (period $\lambda/d=\SI{2.05}{\arcsecond}$), $V\simeq0.94$,
reproducing the curve to $\approx\SI{1.8}{\percent}$ rms [Fig.~3]. The same pipeline applied
to the real-space far-field profile [Fig.~4] gives $d=\SI{9.9}{\micro\meter}$.

\section{Controlled-tilt series and absorber-thickness measurement}
\label{sec:tilt}

\begin{figure}[h]
\centering
\includegraphics[width=0.95\linewidth]{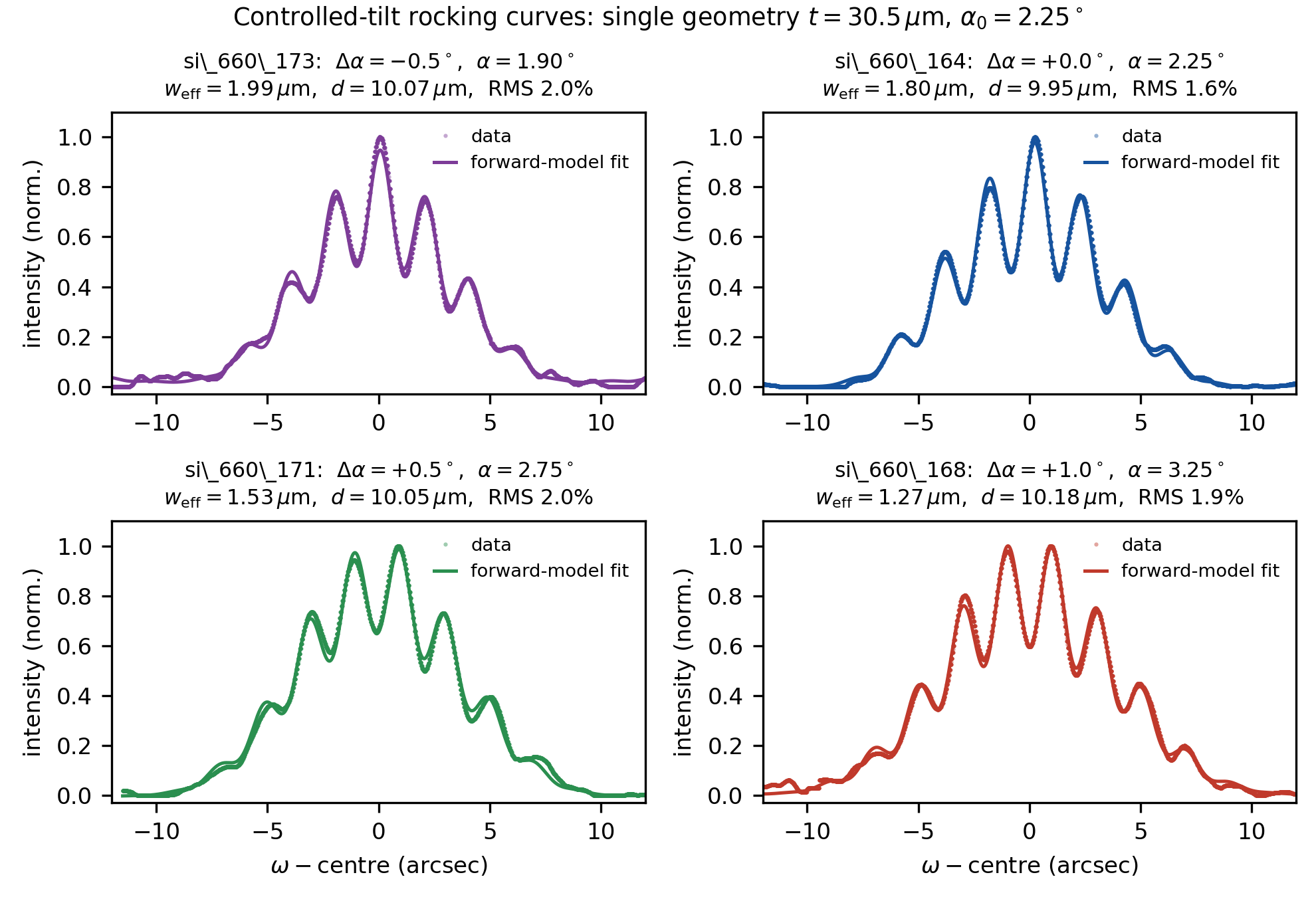}
\caption{The four controlled-tilt rocking curves (measured data, points) with their
soft-edge forward-model fits (lines), computed from the \emph{single} shared geometry
($t=\SI{30.5}{\micro\meter}$, $\alpha_0=2.25^\circ$). The $\Delta\alpha=-0.5^\circ$ run
(si\_660\_173) is held out and its tilt recovered ($\alpha=1.90^\circ$). The single-slit
envelope broadens monotonically with increasing tilt while the fringe spacing---hence
$d\approx\SI{10}{\micro\meter}$---is unchanged.}
\label{fig:tiltcurves}
\end{figure}

\begin{figure}[h]
\centering
\includegraphics[width=0.92\linewidth]{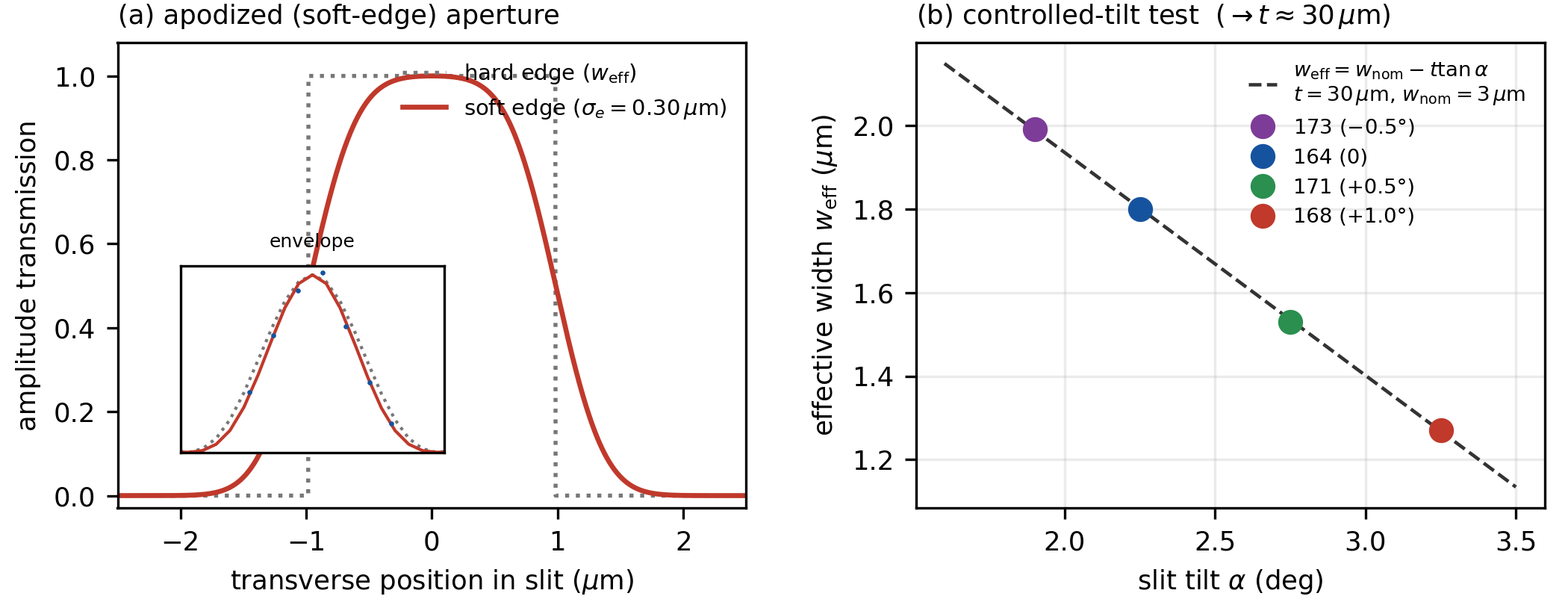}
\caption{Soft-edge aperture model and its controlled-tilt verification.
(a) Effective single-slit amplitude transmission: an ideal hard edge of width
$w_{\rm eff}$ (dotted top-hat) versus the apodized aperture with smoothly graded edges of
width $\sigma_e\approx\SI{0.30}{\micro\meter}$ (solid). Inset: the corresponding far-field
envelope, hard-edge $\mathrm{sinc}^2$ (dotted) versus soft-edge (solid), with the measured
fringe-peak heights (points); the soft edge suppresses the slowly decaying $\mathrm{sinc}^2$
side-lobes and raises the central lobe, matching the data.
(b) Controlled-tilt test: the effective width at the four known tilts (points;
$\Delta\alpha=-0.5^\circ$ to $+1.0^\circ$, Table~\ref{tab:tilt}) falls on the single
projection line $w_{\rm eff}=w_{\rm nom}-t\tan\alpha$ (dashed), whose slope fixes the gold
thickness $t\approx\SI{30}{\micro\meter}$ (with $w_{\rm nom}=\SI{3}{\micro\meter}$).}
\label{fig:soft}
\end{figure}

Because the single-slit envelope encodes $w_{\rm eff}=w_{\rm nom}-t\tan\alpha$, rotating the
\emph{same} slit by known tilt increments $\Delta\alpha$ provides a direct, predictive test
and measures the absorber thickness $t$. Four rocking curves were recorded at
$\Delta\alpha=-0.5^\circ,0,+0.5^\circ,+1.0^\circ$. A \emph{single} geometry
($w_{\rm nom}=\SI{3}{\micro\meter}$, $t=\SI{30.5}{\micro\meter}$, $\alpha_0=2.25^\circ$)
reproduces all four (the $\Delta\alpha=-0.5^\circ$ run is held out and recovered), with the
effective width decreasing monotonically along $w_{\rm eff}=w_{\rm nom}-t\tan\alpha$ and the
pitch $d$ rotation-invariant (Table~\ref{tab:tilt}, Fig.~\ref{fig:soft}(b)). Because $\Delta\alpha$
is known, the slope fixes the otherwise thickness--tilt-degenerate absorber thickness,
$t\approx\SI{30}{\micro\meter}$, consistent with the opacity inferred from the absent
direct-transmission pedestal. The four measured rocking curves and their forward-model
fits are shown individually in Fig.~\ref{fig:tiltcurves}: the single-slit envelope
broadens monotonically with increasing tilt while the fringe spacing (hence $d$) stays fixed.

\begin{table}[h]
\centering
\caption{Controlled-tilt series: single shared geometry ($t=\SI{30.5}{\micro\meter}$,
$\alpha_0=2.25^\circ$).}
\label{tab:tilt}
\begin{ruledtabular}
\begin{tabular}{lccccc}
run & $\Delta\alpha$ & $\alpha$ & $w_{\rm eff}$ (\si{\micro\meter}) & $d$ (\si{\micro\meter}) & rms \\
\hline
si\_660\_173 & $-0.5^\circ$ & $1.90^\circ$ & 1.99 & 10.05 & \SI{1.95}{\percent} \\
si\_660\_164 & $0$          & $2.25^\circ$ & 1.80 & 9.95  & \SI{1.57}{\percent} \\
si\_660\_171 & $+0.5^\circ$ & $2.75^\circ$ & 1.53 & 10.05 & \SI{2.05}{\percent} \\
si\_660\_168 & $+1.0^\circ$ & $3.25^\circ$ & 1.27 & 10.18 & \SI{1.86}{\percent} \\
\end{tabular}
\end{ruledtabular}
\end{table}

